\documentclass{aa}  

\usepackage{graphicx}
\usepackage{txfonts}
\usepackage{lipsum}
\usepackage{subcaption}         
\usepackage{lscape}             
\usepackage{placeins}           
\usepackage{adjustbox}
                                
\newcommand{\iras}{IRAS~18293$-$0941}
\newcommand{\lhaaso}{LHAASO~J1831$-$1007u$^*$}

\begin{document}

   \title{A Galactic microblazar as a potential accelerator \\  of ultra-high-energy particles}


%
%
%

\author{
Josep Mart\'{\i}\inst{1,3}\corrauth{jmarti@ujaen.es}
\and Pedro L. Luque-Escamilla\inst{2,3}\email{peter@ujaen.es}
\and Benito Marcote\inst{4,5}\email{marcote@jive.eu}
\and Leandro Abaroa\inst{6,7}\email{leandroabaroa@gmail.com}
\and Arnau Aguasca-Cabot\inst{8}\email{arnau.aguasca@fqa.ub.edu}
\and Jorge A. Combi\inst{6,7}\email{jorgearielcombi@gmail.com}
\and Gustavo E. Romero\inst{6,7}\email{gustavo.esteban.romero@gmail.com}
\and Josep M. Paredes\inst{8}\email{jmparedes@ub.edu}
\and Federico Garc\'{\i}a\inst{6,7}\email{fgarcia1012002@gmail.com}
\and Federico Fogantini\inst{6}\email{federico.fogantini@gmail.com}
\and Enzo A. Saavedra\inst{9,10}\email{saavedraenz@gmail.com}
\and  Daniel del Ser\inst{11,8}\email{danieldelser@ub.edu}
\and Jakob van den Eijnden\inst{12}\email{a.j.vandeneijnden@uva.nl}
}

\institute{
Departamento de F\`{\i}sica, Escuela Polit\'ecnica Superior de Ja\'en, Universidad de Ja\'en, Campus Las Lagunillas s/n, A3-420, E- 23071 Ja\'en, Spain
\and
Departamento de Ingenier\'{\i}a Mec\'anica y Minera, Escuela Polit\'ecnica Superior de Ja\'en, Universidad de Ja\'en, Campus Las Lagunillas s/n, A3-008, E-23071 Ja\'en, Spain
\and
Grupo de Investigaci\'on FQM-322, Universidad de Ja\'en, Campus Las Lagunillas s/n, A3-065, E-23071 Ja\'en, Spain
\and
Joint Institute for VLBI ERIC, Oude Hoogeveensedijk 4, 7991 PD, Dwingeloo, The Netherlands 
\and
ASTRON, Netherlands Institute for Radio Astronomy, Oude Hoogeveensedijk 4, 7991 PD, Dwingeloo, The Netherlands 
\and
Instituto Argentino de Radioastronom\'{\i}a, (CCT La Plata, CONICET; CICPBA; UNLP), C.C.5, (1894), Villa Elisa, Argentina 
\and
Facultad de Ciencias Astron\'omicas y Geof\'{\i}sicas, Universidad Nacional de La Plata, B1900FWA La Plata, Argentina 
\and
Departament de Física Qu\`antica i Astrof\'{\i}sica, Institut de Ci\`encies del Cosmos, Universitat de Barcelona, IEEC-UB, Martí\'{\i} i Franqu\' es 1, E-08028 Barcelona, Spain
\and
Instituto de Astrof\'{\i}sica de Canarias (IAC), Calle Vía L\'actea, s/n, E-38205 La Laguna, Tenerife, Spain
\and
Departamento de Astrof\'{\i}sica, Universidad de La Laguna, E-38205 La Laguna, Tenerife, Spain
\and
Observatori Fabra, Reial Acad\`emia de Ci\`encies i Arts de Barcelona, Rambla dels Estudis, 115, E-08002 Barcelona, Spain
\and
Anton Pannekoek Institute for Astronomy, Universiteit van Amsterdam, Science Park 904, NL-1098, XH, Amsterdam, the Netherlands
}

   \date{Received May XX, 2026}

\abstract{
Persistent jets from X-ray binaries which are aligned very close to the line of sight could be considered to be Galactic
equivalents of blazars, or 'microblazars'. They are also expected to power gamma-ray sources.}
{We intend to assess a  
serious candidate apparently fulfilling  
many of the requirements to be considered a genuine member of this class: \iras.  }
{An intense multi-wavelength observational and theoretical study has been carried out on our proposed candidate source.}
{With photometric and spectroscopic properties typical of a binary star, this system exhibits clear collimated and one-sided radio emission matching the effects of relativistic motion along a reduced ejection angle. 
Only fast variability is not observed possibly smoothed by a dense circumstellar envelope. 
A physical scenario is consistently modeled that also gives credibility to its likely connection with 
\lhaaso, an ultra-high-energy source in its immediate vicinity.
}
{Our reported identification not only helps to fill a gap in Galactic taxonomy, but also potentially strengthens  the role of the microblazar and microquasar families in our understanding of the most energetic Milky Way phenomena.}

   
   \maketitle
   \nolinenumbers

\section{Introduction}

Galactic microblazars—microquasars whose relativistic jets are oriented close to the observer’s line of sight—have long been predicted 
\citep{ref01,ref02,ref03},  yet firmly established examples have proven difficult to identify in the Galaxy. Candidates relied mainly on indirect signatures of relativistic boosting \citep{ref04,ref05,ref06}  and did not withstand detailed multi-wavelength scrutiny \citep{ref07,ref08,ref09}. 
 Very recently,  the black hole X-ray binary
 \object{4U 1543$-$47} has been  reported to develop a one-sided high-Lorentz factor jet inclined less than $27^{\circ}$ to the line of sight
 \citep{Zhang2026Jets}. 
 However, this occurred during a transient eruptive event and such behavior therefore lacks long-term persistence. 

In this work, we present \iras\ as a compelling Galactic microblazar candidate. The multi-wavelength observations reported in the following sections reveal a binary system launching a persistent, highly asymmetric jet viewed at a small inclination angle. Very Long Baseline Interferometry combined with {\it Gaia} astrometry firmly associates the radio emission with the stellar system, excluding an extragalactic origin, while independent geometric constraints from optical photometry indicate a nearly face-on orbital configuration, consistent with strong Doppler boosting. Infrared and X-ray observations further show that the jet is actively interacting with a dense and dusty environment, providing direct evidence of jet-driven feedback. At larger scales, deep radio imaging reveals a hotspot–like structure aligned with the jet axis, suggestive of a jet termination region. We show that the power and geometry of \iras\ make it a viable accelerator of 
very-high-energy 
(VHE, $E> 100$ GeV) particles and a plausible contributor to the ultra-high-energy 
(UHE, $E>100$ TeV) 
gamma-ray source \lhaaso\ \citep{ref10}.  Together, these results provide  
a robust observational realization of a Galactic microblazar and highlight their potential role as efficient particle accelerators and agents of kinetic feedback in the Milky Way.


\section{Identification of \iras\ as a microblazar candidate and extensive follow-up}

\begin{figure}
   \centering
   \includegraphics[angle=-0,width=8.1cm]{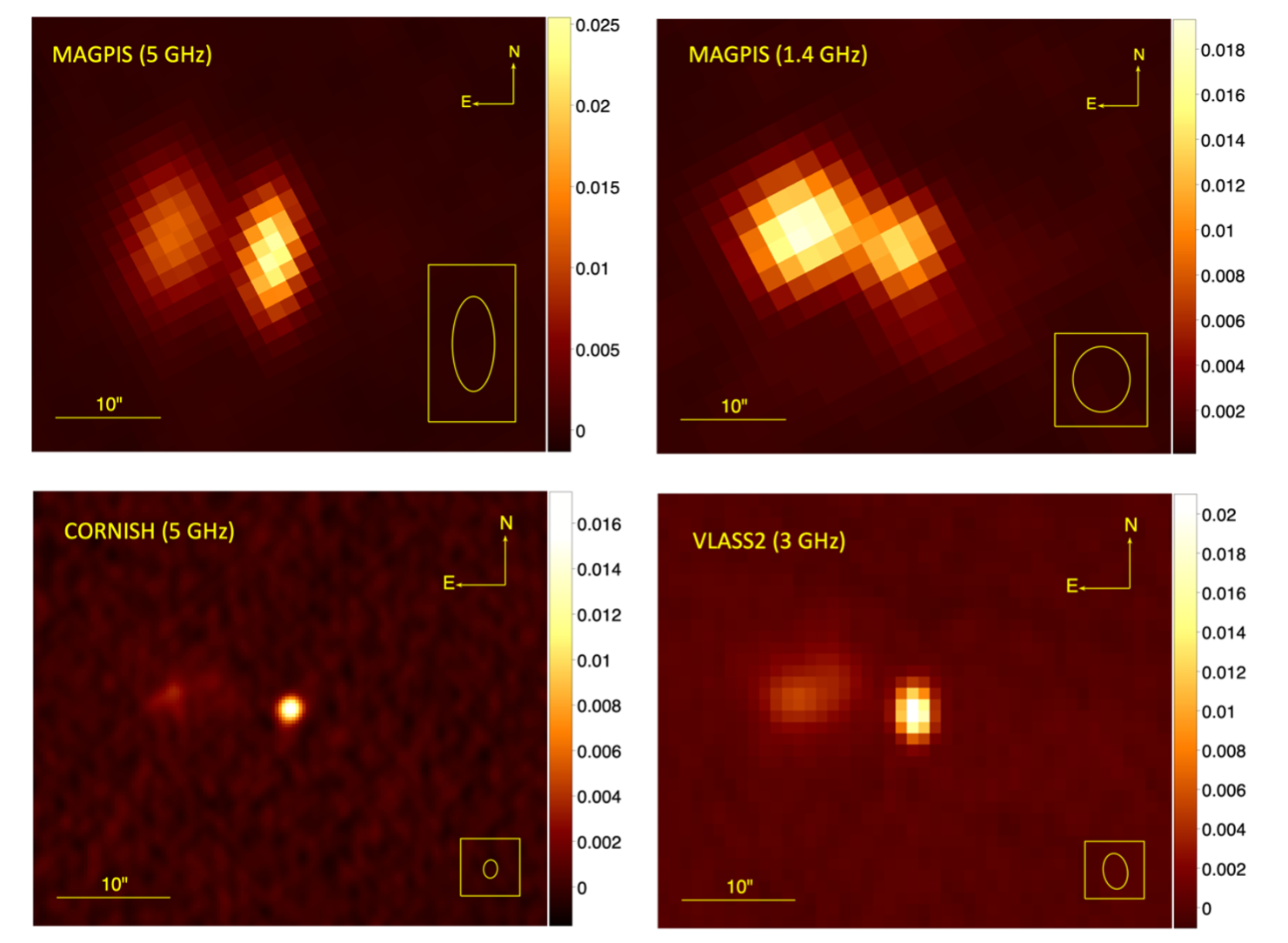}
      \caption{
Radio maps of \iras\ obtained from the MAGPIS \citep{ref13}, CORNISH \citep{ref14} and VLASS \citep{ref15} surveys. In all panels the source appearance is that of a compact core plus a one-sided Eastern lobe. These features are persistent over the different years of the respective survey data acquisition. The observing frequency, brightness scale (expressed in Jy beam$^{-1}$), orientation, scale bar and point spread function are plotted in each panel.
                   }
         \label{S1}
   \end{figure}

\iras\ was originally proposed as a high-mass X-ray binary system based on tentative photometric fits \citep{ref11}, 
but remained poorly studied owing to extreme extinction despite its striking H$\alpha$ emission and radio signatures. 
The source emerged independently to us from a multi-wavelength cross-identification strategy  \citep{ref12} when  focused on catalogued reddened luminous stars.
Soon after, the first suspicion about the possible microblazar nature of \iras\ came from its resolved one-sided appearance 
(see Fig. \ref{S1}) consistently present over the years in different surveys at cm wavelengths \citep{ref13,ref14,ref15}.
Therefore we were clearly facing a persistent phenomenon instead of a transient short-lived feature created during a transient event.
This agrees with the fact that no outburst alert has been ever reported for this source. 

\begin{figure*}
   \centering
   \includegraphics[angle=-0,width=17.95cm]{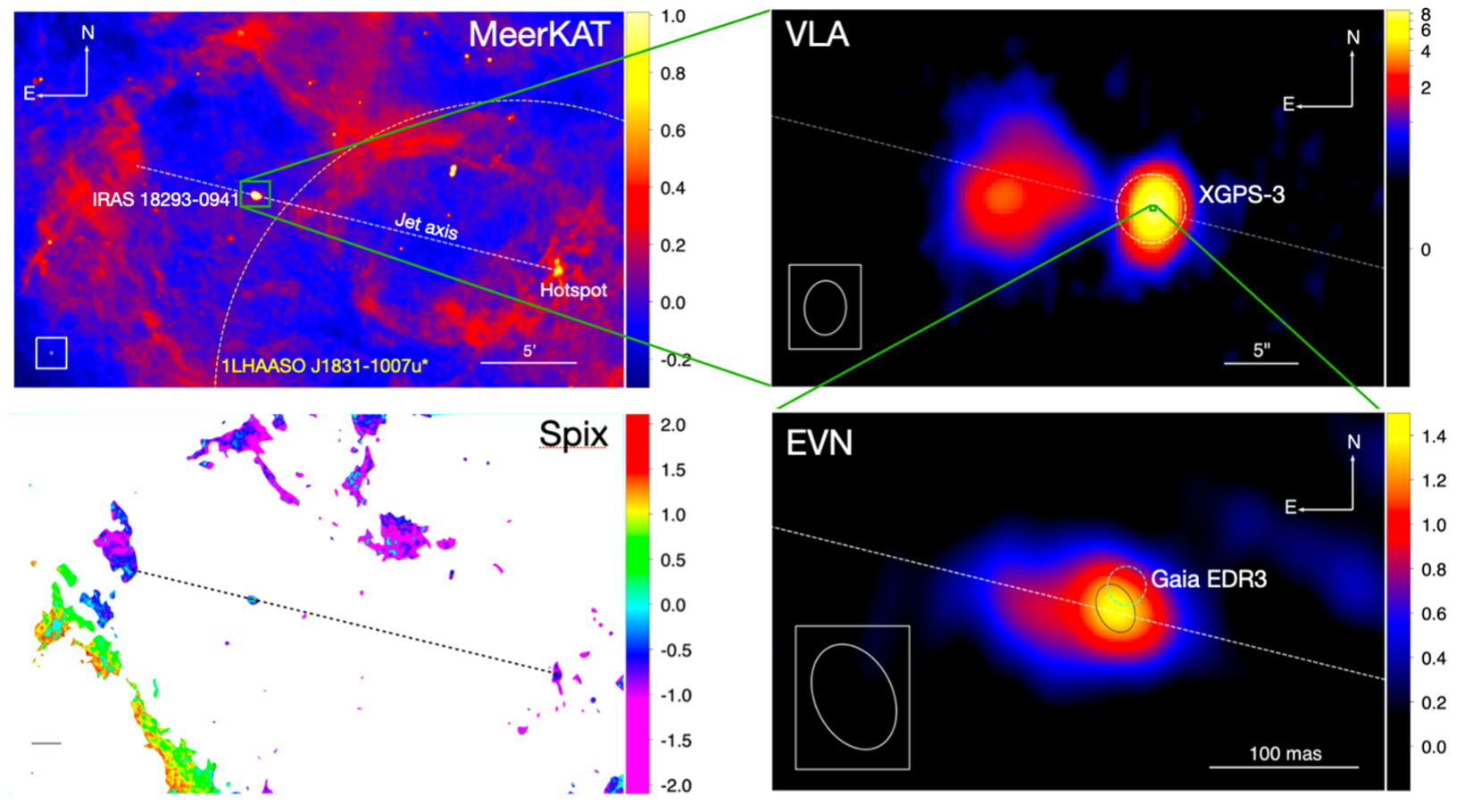}
      \caption{
   Radio view of \iras\  across different spatial scales. {\bf Top right.} Arcsecond-resolution VLA map computed from the combination of the observing runs listed in
    Table \ref{TS1}. It shows a compact and flat-spectrum core from which a one-sided jet emanates powering an arcsecond extended radio lobe. The yellow dashed circle corresponds to the 90\% confidence location of the X-ray source XGPS-3 \citep{ref11}. {\bf Bottom right.} Sub-arcsecond EVN radio map resolving the VLA core into an also one-sided jet aligned with the larger-scale lobe structure. The cyan and black ellipses mark the 95\% confidence regions where the {\it Gaia} DR3 position 
    \citep{ref16} of \iras\  and the radio photocenter are expected (given astrometric uncertainties and proper motion correction), respectively. 
    {\bf Top left.} Arcminute-scale from the MeerKAT radio survey \citep{ref20} showing arc-shaped extended emission around the IRAS source and a hotspot feature aligned with the lobe and jet axis. {\bf Bottom-left.} Dimensionless spectral index map available thanks to the large MeerKAT bandwidth. Brightness scale is provided for the rest of panels in mJy beam$^{-1}$, with beam sizes displayed in all bottom left corners. Dashed white lines indicate the common jet direction across more than four orders of magnitude in angular size along a position angle of  $256^{\circ}$ from North to East.
                   }
         \label{F1}
   \end{figure*}

\begin{figure*}
   \centering
   \includegraphics[angle=-0,width=18.0cm]{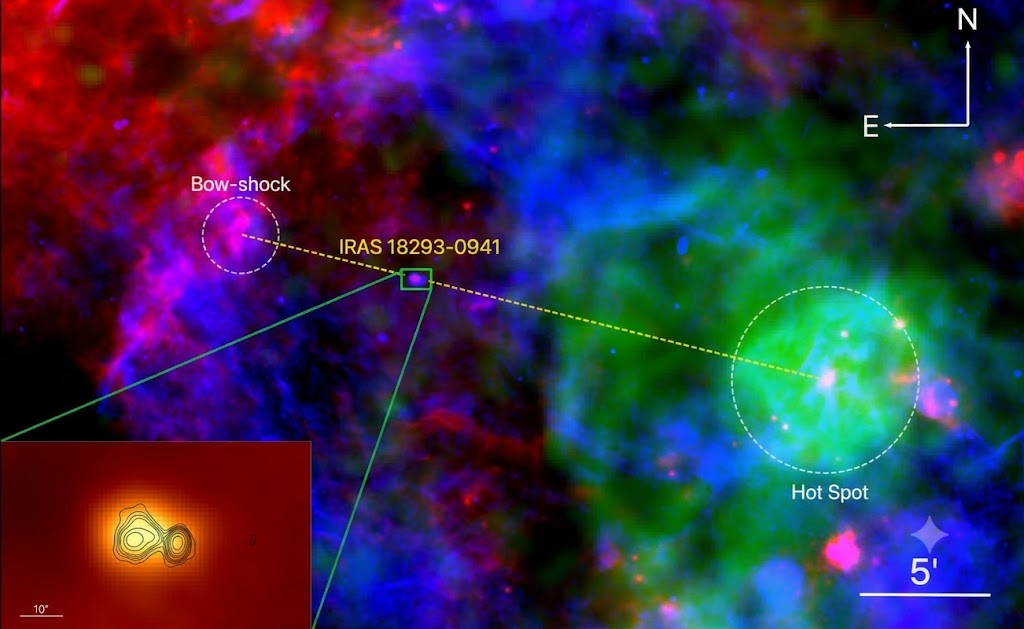}
      \caption{
      Jet–environment interaction in \iras\ on multiple spatial scales. 
      Tri-chromatic composition with red, green and blue layers representing the Herschel 70 $\mu$m emission \citep{ref19}
       tracing warm dust, the H$\alpha$ emission \citep{ref24} tracing shock-ionized gas and MeerKAT radio emission \citep{ref20} 
       mostly non-thermal, respectively. The two jets exhibit a pronounced asymmetry. The Western receding jet terminates in a confined non-thermal hotspot coincident with enhanced infrared emission and extended H$\alpha$ nebulosity, revealing a strong bow shock likely produced by the impact of the relativistic jet on a dense molecular cloud. In contrast, the Eastern approaching jet expands into a broader and more diffuse radio structure, with no associated optical emission, consistent with propagation into a lower-density medium. The position of the stellar system is marked with a small rectangle zoomed at the bottom left corner. The close spatial correspondence between the VLA radio lobe (black contours) and the dust emission demonstrates that the outflow interacts with a dense environment already at small distances from the central binary. Angular scale bars are shown for each panel and the compass direction applies to both.
               }
         \label{F2}
   \end{figure*}

\begin{figure}
   \centering
   \includegraphics[angle=-0,width=\columnwidth]{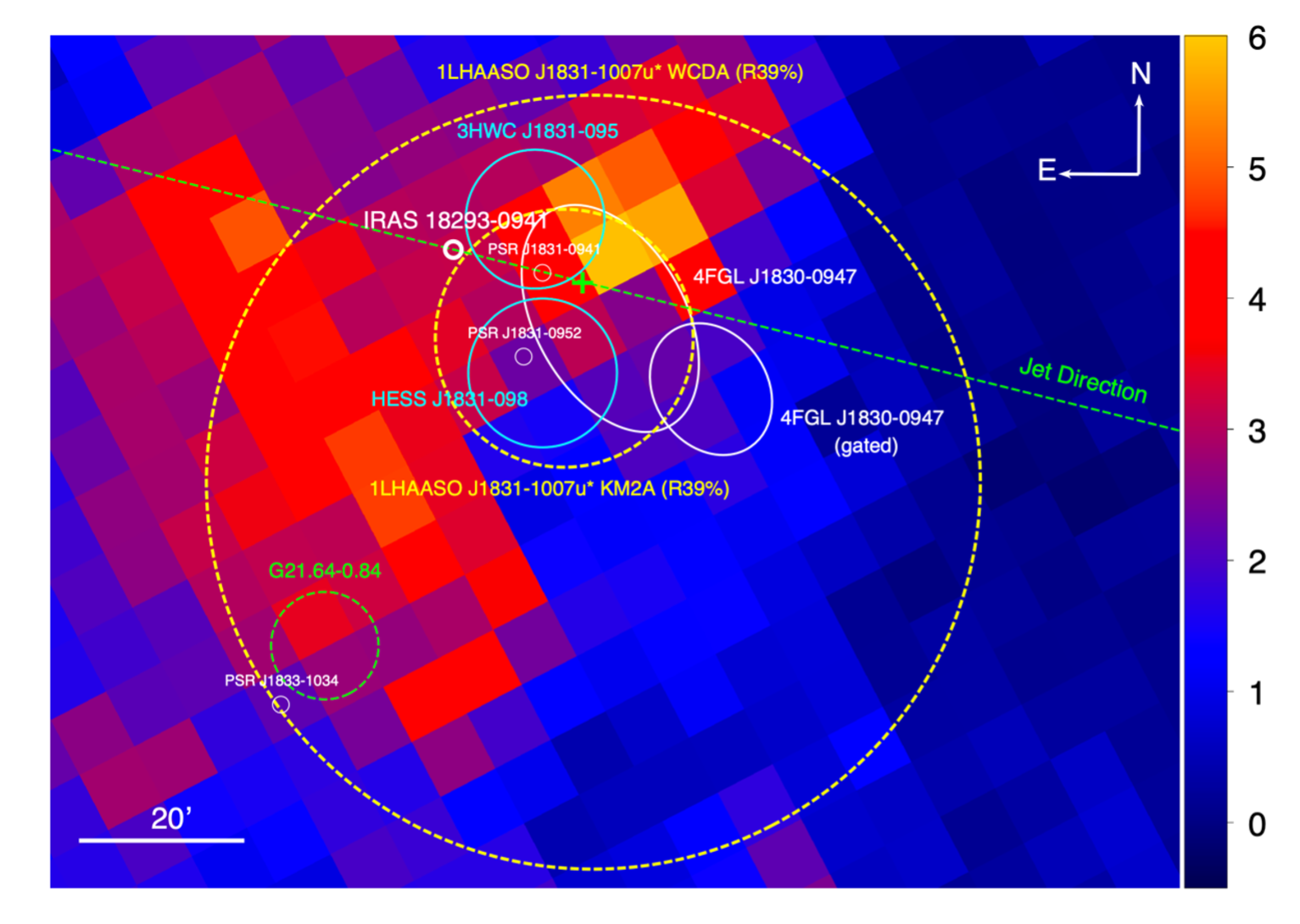}
      \caption{
 Molecular cloud environment of \iras. Map of survey CO emission \citep{ref42} showing the position of \iras\ relative to HE, VHE and UHE gamma-ray sources in the field. The background is dominated by the molecular cloud (21.97, $-$0.29) \citep{ref25}. Relevant pulsars in the field are indicated as small white circles, while the thick one is reserved for the IRAS source. The locations of Fermi-LAT, H.E.S.S., HAWC, and LHAASO sources, taken from their respective catalogues, are shown as ellipses. The reprocessed Fermi-LAT location after pulsar gating is also plotted. For LHAASO, both WCDA and KM2A, the radii of 39\% source flux are used due to extended nature. Color codes are chosen to match labels and source positions. The supernova remnant G21.64$-$0.84 in the region outskirts, is also displayed. The spatial coincidence between the dense jet-cloud interaction zone and the LHAASO uncertainty region motivates a physical association. Position angle of the jet direction is indicated by the dashed green line, while a small green cross marks its proposed hotspot. The vertical scale represents CO brightness temperature in Kelvin. North is up and East is left, with the horizontal bar indicating the angular scale.
              }
         \label{F3}
   \end{figure}

\begin{figure}
   \centering
   \includegraphics[angle=-0,width=\columnwidth]{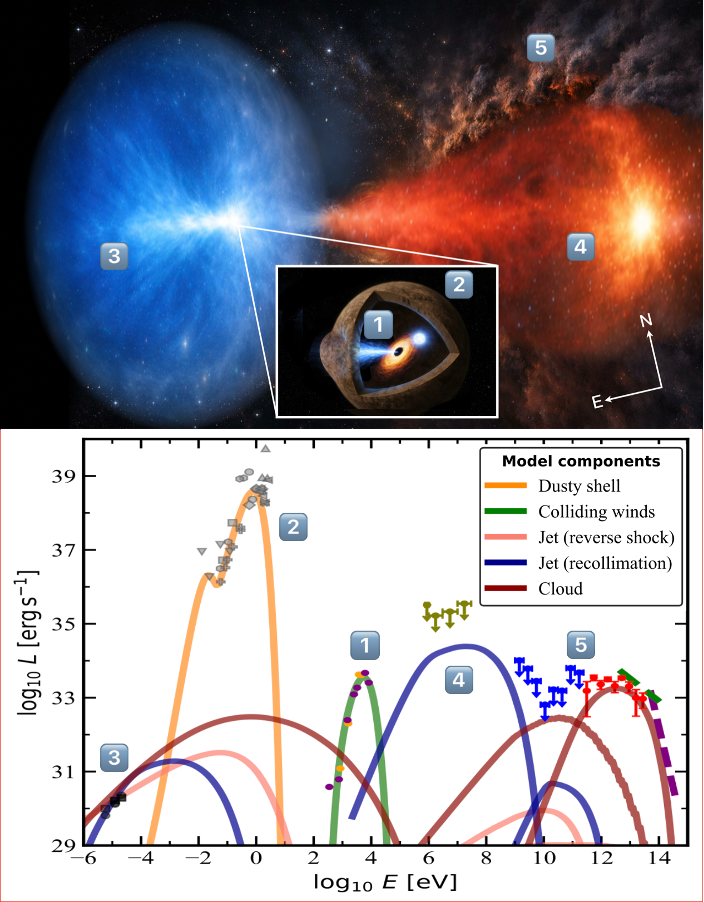}
      \caption{      
      Multi-zone scenario for \iras\ and modelled broadband emission. 
      {\bf (Top)} Artistic view of the \iras\ environment. The system is interpreted as a microblazar powered by a black hole accreting at super-Eddington rate from a donor star through an accretion disc whose optically thick winds collide with the stellar wind (1). A dense circumbinary dusty shell surrounds the system (2)
       and is distorted by the approaching  (blue) jet (3), which eventually breaks out and expands into the lower-density ambient medium, inflating a cocoon. The receding (red) jet (4) propagates toward a dense molecular cloud (5), where it produces a radio hotspot and injects accelerated cosmic rays into the cloud, giving rise to high-energy emission through hadronic processes. 
    {\bf   (Bottom)} Multi-zone broadband, dereddened, non-simultaneous SED reproduced by our model. Labels indicate the dominant emission regions. Observational data points combine archival measurements and dedicated observations, covering radio frequencies (from the compact core to arcminute-scale jets; EVN, VLA, CORNISH, MAGPIS, VLASS), optical and infrared bands (Pan-STARRS, {\it Gaia}, SDSS, 2MASS, DENIS, WISE, MSX, AKARI, IRAS, VST/OmegaCAM), X-rays (XMM–Newton, Chandra, Swift), and extending to the high- and very-high-energy gamma-ray domains (COMPTEL, HESS, HAWC, LHAASO).
                  }
         \label{F4}
   \end{figure}

After recognizing the potential interest of this target, we started an intensive follow-up both observational and theoretical.
Given the large amount of data collected,
we present our results using an Appendix structure to alleviate the reading effort.
 The aspects covered include: 
 improved radio mapping  (Appendices  \ref{mejora}, \ref{espectroradio}, \ref{ratio}, and \ref{VLBI}),
 optical  monitorings (Appendices  \ref{caha} and \ref{photo}), associated large-scale features (Appendices  \ref{hospot} and \ref{spix}),
 possible shock tracers (Appendix \ref{shocks}), jet-dust interaction (Appendix \ref{Herchel}), kinematic distance (Appendix \ref{dist}),
 X-ray and gamma-ray analysis (Appendices  \ref{xray} and \ref{arnau}), and theoretical modelling 
 of the Spectral Energy Distribution (SED, Appendix \ref{model}).  
 The most relevant findings are those displayed in main text Figs. \ref{F1}, \ref{F2}, \ref{F3} and \ref{F4}, as discussed below and in their respective Appendices .
 We will refer to these different follow-up results as needed while  justifying and converging towards our final conclusions.

\section{Discussion}

Relativistic jets launched by stellar-mass compact objects efficiently transport energy into the interstellar medium and produce non-thermal emission across the electromagnetic spectrum. In favorable geometries, Doppler boosting strongly amplifies their observed emission, potentially giving rise to microblazars  or Galactic analogues of extragalactic blazars. For randomly oriented jets, such configurations are expected to be rare, consistent with the scarcity of 
 confirmed examples to date. 

As evidenced from our \iras\ follow-up,
 archival Very Large Array (VLA) 6 cm observations and a dedicated European VLBI Network (EVN) run at 18 cm resolve the source into an elongated, flat-spectrum central component aligned with the larger-scale and non-thermal radio lobe to the East (see Fig. \ref{F1} right panels and radio spectra in Fig. \ref{S2}). Such morphology is characteristic of a relativistic jet viewed at small inclination, with strong Doppler boosting suppressing the opposite counterjet
 (details in Appendices  \ref{mejora}, \ref{espectroradio}, \ref{ratio} and \ref{VLBI}). 
 
 Crucially, the milliarcsecond precision (VLBI) position is consistent 
 within uncertainties with the high-precision {\it Gaia} DR3 astrometric solution for the optical counterpart \citep{ref16}, firmly associating the radio jet with the stellar system itself and excluding a background active galactic nucleus (Appendix \ref{VLBI})
 The absence of any measurable cosmological redshift in the stellar-like optical spectrum (Appendix \ref{caha}) further confirms a Galactic origin.
 
In contrast to previously proposed systems $-$most notably V4641 Sgr \citep{ref17} $-$ \iras\ exhibits both a steady jet morphology and independent geometric constraints on its orientation, thereby closely mimicking the appearance criteria that would be required to be classified as a blazar \citep{ref18}. 
This striking resemblance strongly reinforces our identification as a Galactic microblazar candidate since we are certainly dealing with a Milky Way object. The pronounced asymmetry between the jet and counterjet implies strong relativistic Doppler boosting. Constraints on the jet-to-counterjet brightness ratio require a relativistic outflow oriented not far from the line of sight (Appendix \ref{ratio}). 
Independent geometric information is provided by long-term optical $I$-band photometric monitoring. Periodogram analyses reveal a dominant modulation with a period of $11.38 \pm 0.01$ day, consistent with orbital variability in a binary system  
(see Figs. \ref{S5} and \ref{S6}). 
 The extremely low amplitude of the ellipsoidal modulation and the absence of eclipses indicate a low orbital inclination, likely  $\leq 30^{\circ}$. Assuming ejection perpendicular to the orbital plane, this is in agreement with the jet orientation inferred from radio observations. Together, these constraints strongly support a near face-on geometry characteristic of a microblazar (Appendix \ref{photo}).

Unanticipatedly, no significant optical flickering is detected on intra-hour timescales within the sensitivity of the data ($\sim 0.01$ mag). This can be reconciled with a microblazar interpretation, as the optical emission in the accessible optical $I$-band is likely dominated by the stellar component and radiation reprocessed in a dense, dusty circumstellar envelope, which can suppress short-timescale variability originating in the inner accretion flow or jet. 
Infrared and X-ray observations corroborate this picture (Appendices  \ref{Herchel} and \ref{xray}). Although spectral classification is hampered by reddening, the strong emission lines and P-Cygni profiles point to an expanding envelope around a likely early-type companion (Appendix \ref{caha}). This is strengthened by the  infrared SED  that shows a pronounced excess consistent with substantial circumstellar dust reprocessing, while the steady, unpulsed X-ray counterpart, coincident with the radio core and optical star, exhibits a shocked-gas spectrum with significant absorption. 

Direct evidence of jet-environment interaction is provided by archival far-infrared observations (Fig. \ref{F2} and Appendix \ref{Herchel})
 from the Herschel Space Observatory \citep{ref19}. At 70 $\mu$m, warm dust emission displays a shell-like morphology closely aligned with the radio jet axis at arcsecond angular scales (Fig. \ref{F2} inset), indicating that the relativistic outflow is sweeping up and compressing circumstellar material in the immediate surroundings of the binary. The tight spatial correspondence between radio and far-infrared emission reveals ongoing jet-driven feedback already within the dense envelope of the system.

At larger spatial scales (arcminutes, Appendix \ref{hospot}), MeerKAT survey imaging \citep{ref20} is suggestive of  
a limb-brightened, 
jet-inflated radio cavity aligned with the compact jet axis (Fig. \ref{F1}, left panels), with emission predominantly tracing its outer boundaries rather than filling its interior, reminiscent of large-scale jet-blown structures observed among others in Galactic microquasars such as Cygnus X-1 \citep{ref21}, GRS 1915+105 \citep{ref22}, 
 SS 433 \citep{dubner1998ss433}, and Cir X-1 \citep{tudose2006circinus}.
On the receding-jet side, to the West, a localized surface-brightness enhancement that we tentatively interpret as 
a distinct terminal interaction region. This apparent hotspot feature is spatially coincident with compact H$\alpha$ emission detected in the VST Photometric H$\alpha$ Survey of the Southern Galactic Plane (VPHAS+) \citep{ref23} 
and the Northern Sky Narrowband Survey (NSNS) \citep{ref24}, pointing to shock-excited ionized gas (Fig. \ref{F2}). 
The MeerKAT radio spectral index in this region is non-thermal ($\alpha = -0.7 \pm 0.1$), 
lending support to  
ongoing particle acceleration at the proposed termination shock (Appendix \ref{spix}). 
Surrounding this compact acceleration site, the radio emission resolves into filamentary structures that closely trace the H$\alpha$ nebula. Their tight morphological coupling to the synchrotron hotspot also supports shock-driven ionization powered by the dissipation of jet kinetic energy (Appendix \ref{shocks}).
 Co-spatial 70 $\mu$m emission indicates the presence of compressed and heated dust within the post-shock region, consistent with a radiative shock propagating into a dense ambient medium (Appendix \ref{Herchel}). The compact morphology and strong confinement of the radio emission indicate that the jet remains efficiently collimated up to its terminal interaction with a dense molecular structure, where its kinetic energy is deposited into a relatively small volume (see again Fig. \ref{F1} left panels).

In addition to the localized hotspot, the inspected H$\alpha$ surveys further reveal faint extended line emission co-spatial with portions of the radio-brightened cavity edges (Fig. \ref{F2}). The spatial correspondence between optical H$\alpha$ emission and the cocoon edge supports the presence of large-scale shock fronts driven by the expanding jet-inflated cavity. 
Conversely, no compact terminal hotspot is detected by MeerKAT on the Eastern approaching-jet side (Fig. \ref{F1} top left panel). The radio emission instead delineates a more diffuse and spatially larger limb-brightened cavity, with only mild surface-brightness enhancement along its outer boundary. This structure reaches a smaller projected distance along the jet axis than the receding side, while exhibiting a significantly larger transverse extent. Such morphology indicates that, in a lower-density environment, reduced external pressure permits enhanced lateral expansion of the flow. The resulting partial de-collimation distributes the jet momentum over a broader cross-section, inflating a wide cocoon whose axial advance is comparatively limited. Although far-infrared emission at 70 $\mu$m is also present in this direction, it predominantly traces the limb-brightened edges of the radio cavity rather than the jet axis, and no localized optical line emission zone is detected. However, a less prominent bow-shock region does appear to exist aligned with the extrapolated Eastern jet axis (Fig. \ref{F2}). With a non-thermal radio spectral index, it is spatially coincident with a 70 $\mu$m curved arcminute feature. This suggests a weaker or more distributed interaction between the jet and the ambient medium. The absence of strong optical shock tracers and the lower degree of radio confinement indicate propagation into a significantly lower-density environment. Together, these observations reveal a pronounced environmental density gradient along the jet axis, which naturally explains the marked asymmetry in collimation, projected extent, and dust distribution. This is consistent with the system residing at the interface between a dense molecular cloud and the more diffuse interstellar medium.

The presence of dense gas at the location of the receding-jet termination is independently supported by CO line survey data \citep{ref25}, 
which identify a molecular cloud at 3.6 kpc towards the \iras\  direction. It is therefore conceivable that observed interaction evidences (cavity and hotspot) arise from the jet impact on it. This renders the cloud kinematic distance value as a plausible estimate for the IRAS source as well (Appendix \ref{dist}). 
X-ray emission, mostly attributed to collision of stellar and accretion disk winds, implies at this distance a 0.5-10 keV luminosity of order $\sim 10^{34}$ erg s$^{-1}$
(Appendix \ref{xray}).

The receding jet impacts then directly onto the dense cloud, naturally acting as an efficient hadronic beam dump for high-energy particles accelerated in the jet. In this context, the positional coincidence of the large-scale jet interaction region also with the UHE 
 gamma-ray source 
LHAASO J1831$-$1007u$^*$ \citep{ref10}  becomes easily explained (Fig. \ref{F3}). 
The dense molecular material provides an optimal target for hadronic interactions of relativistic protons escaping from the jet termination region, leading to the production of UHE gamma rays. Although LHAASO J1831$-$1007u$^*$ has   been tentatively associated with other objects previously reported in the field \citep{ref26}, 
the jet geometry, the clear evidence for large-scale termination shocks, and the strong asymmetry in environmental density together motivate an alternative scenario in which \iras\ is the dominant contributor to the observed UHE emission (Appendix \ref{arnau}).

To assess the internal consistency of this scenario, we constructed a broadband SED model that simultaneously accounts for the radio, infrared, X-ray and gamma-ray emission components (Appendix \ref{model}). 
The inferred jet power and sustained activity imply that \iras\ is capable of accelerating particles to VHE 
and potentially UHE energies, in line with recent studies linking microquasars to Galactic PeVatrons \citep{ref17,ref27, ref28, ref29, ref30, ref31}. The system also lies within a region hosting previously reported GeV-TeV emission (Fig. \ref{F3}) detected by other instruments \citep{ref32,ref33,ref34}, suggesting that particle acceleration and escape may have operated over extended timescales.

The resulting SED fit in Fig. \ref{F4} (bottom panel) demonstrates that a single physical framework reproduces the multi-wavelength phenomenology of the \iras\ system, as sketched in Fig. \ref{F4} (top panel). Synchrotron emission dominates the radio band, dust reprocessing produces the infrared excess, shocked winds account for the X-rays, and escaping high-energy particles illuminate the nearby molecular cloud. The region responsible for the highest-energy emission may extend well beyond the presently observed radio structures, suggesting that the UHE phenomenology could also reflect past phases of enhanced jet activity \citep{ref35}.

\section{Conclusions}

The main conclusions of our work are summarized as follows:

\begin{enumerate}

\item The presented observations and the proposed  theoretical scenario, together,  establish \iras\ as 
a compelling Galactic microblazar candidate.
 It exhibits a persistent relativistic jet consistent with a reduced inclination angle, with independent geometric support of its orientation from optical photometry.
 This system further illustrates how expected properties, such as rapid variability, can be masked by a dense environment of gas and dust.
  
 \item Suggestive evidence of ejecta interaction with a dense local medium on multiple spatial scales 
 has been also provided. Interestingly, 
 a highly credible non-thermal hot spot feature has been identified tens of parsecs away in the receding jet direction that will deserve further study.

\item  Lastly, \iras\ is introduced as a key component in a region characterized by complex gamma-ray emission, suggesting a physical connection with the UHE source LHAASO J1831$-$1007u$^*$. In this framework, the PeV emission is naturally explained by charged-pion decay resulting from the interaction between relativistic jets and nearby molecular clouds. This scenario successfully reproduces the observed SED across the entire spectrum, from radio to gamma-rays.

\end{enumerate}

\begin{acknowledgements}

The National Radio Astronomy Observatory Karl G. Jansky Very Large Array is a facility of the U.S. National Science Foundation operated under cooperative agreement by Associated Universities, Inc. The European VLBI Network is a joint facility of independent European, African, Asian, and North American radio astronomy institutes. Scientific results from data presented in this publication are derived from the following EVN project code: EM178.  Based on observations collected at the Centro Astron\'omico Hispano en Andaluc\'{\i}a (CAHA) at Calar Alto, proposal 25A-2.2-001, operated jointly by Junta de Andaluc\'{\i}a and Consejo Superior de Investigaciones Cient\'{\i}ficas (IAA-CSIC). This work is partially based on observations obtained with the Zwicky Transient Facility, funded by the NSF and a consortium of institutions. The Joan Or\'o Telescope at the Montsec Observatory (OdM) is owned by the Catalan Government and operated by the Institute of Space Studies of Catalonia (IEEC). This work is partially based on observations obtained with XMM-Newton, an ESA science mission with instruments and contributions directly funded by ESA Member States and NASA, and retrieved from the XMM-Newton Science Archive (XSA). This research has made use of data obtained from the Chandra Data Archive provided by the Chandra X-ray Center (CXC). The use of archival Fermi-LAT data provided by the Fermi Science Support Center is also acknowledged. 

This work was supported by the State Agency for Research of the Spanish Ministry for Science and Innovation under grants PID2022-136828NB-C42/MCIN/AEI/ 10.13039/501100011033/ERDF/EU, PID2022-136828NB-C41/MCIN /AEI/10.13039/501100011033/ERDF/EU, PID2021-124879NB-I00 and PID2024-161863NB-I00, through the “Unit of Excellence María de Maeztu” award to the Institute of Cosmos Sciences (CEX2019-000918-M and CEX2024-001451-M). Also supported by the E.U. Horizon 2020 research and innovation programme under projects `EuroFlash’ (Grant agreement No. 101098079), ‘OCEANS’ (Grant agreement No. 101183150), and ‘MeerSHOCKS’ (ERC Marie Skłodowska-Curie Grant agreement No 101148693).

\end{acknowledgements}

%

\bibliography{references}

\begin{appendix}

\twocolumn

\section{An improved radio image} 
\label{mejora}

To obtain a better view of \iras, we searched the National Radio Astronomy Observatory (NRAO) public archives for historical Very Large Array (VLA) observations of this target and identified several datasets providing high-quality imaging at similar wavelengths. The observations best suited to characterize the source morphology were obtained at 6 cm in the array B and C configurations, which offer a good compromise between angular resolution and sensitivity to compact and moderately extended emission (Table \ref{TS1}). These runs included different scans over 3C286 and the pair 1804+010 and 1820$-$254 that were observed as amplitude and phase calibrators, respectively. All datasets were re-calibrated and imaged using standard procedures within the AIPS software package. The oldest observations (1990 and 2004), originally recorded in the B1950 reference frame, were converted to J2000 using the task UVFIX. Individual runs were first imaged separately with IMAGR to assess data quality and subsequently concatenated using DBCON, after shifting all phase centers to a common reference position. Calibrated visibilities for the latest 2006 run could be directly retrieved as they were part of the CORNISH radio survey \citep{ref14}. Final imaging was performed with IMAGR. 

The resulting 6 cm combined map (Fig. \ref{F1}, top right panel) reveals a compact radio core coincident with the optical position of \iras, accompanied by a one-sided elongated structure extending up to $\simeq10^{\prime\prime}$. The pronounced asymmetry and collimated morphology strongly support a jet origin.

\begin{table}
\caption{Log of VLA archive 6 cm observations(*)  used for imaging purposes}
\label{TS1}
\centering
\begin{tabular}{c c c c c}
\hline\hline                 
 Project & Epoch & VLA    &  Bandwidth &       Number of \\
  Code    &            &     Config. &  (MHz)      &  Visibilities \\
 \hline
AB573   & 1990/12/07 & C &   50 &  5850\\
              & 1990/12/09 & C &   50 &  5850\\
AB1112  & 2004/04/28 & C &   50 &  7475 \\
AH884(**) & 2006/07/12 & 	B &  25 & 14553\\
\hline
\end{tabular}
\tablefoot{\\
(*) Central frequency of 4.885 GHz with two IF pairs. \\
(**) $uv$ data from CORNISH survey \citep{ref14}.}
\end{table}

\section{Radio spectral properties of the core and arcsecond extended component} 
\label{espectroradio}

\begin{figure}
   \centering
   \includegraphics[angle=-0,width=8.0cm]{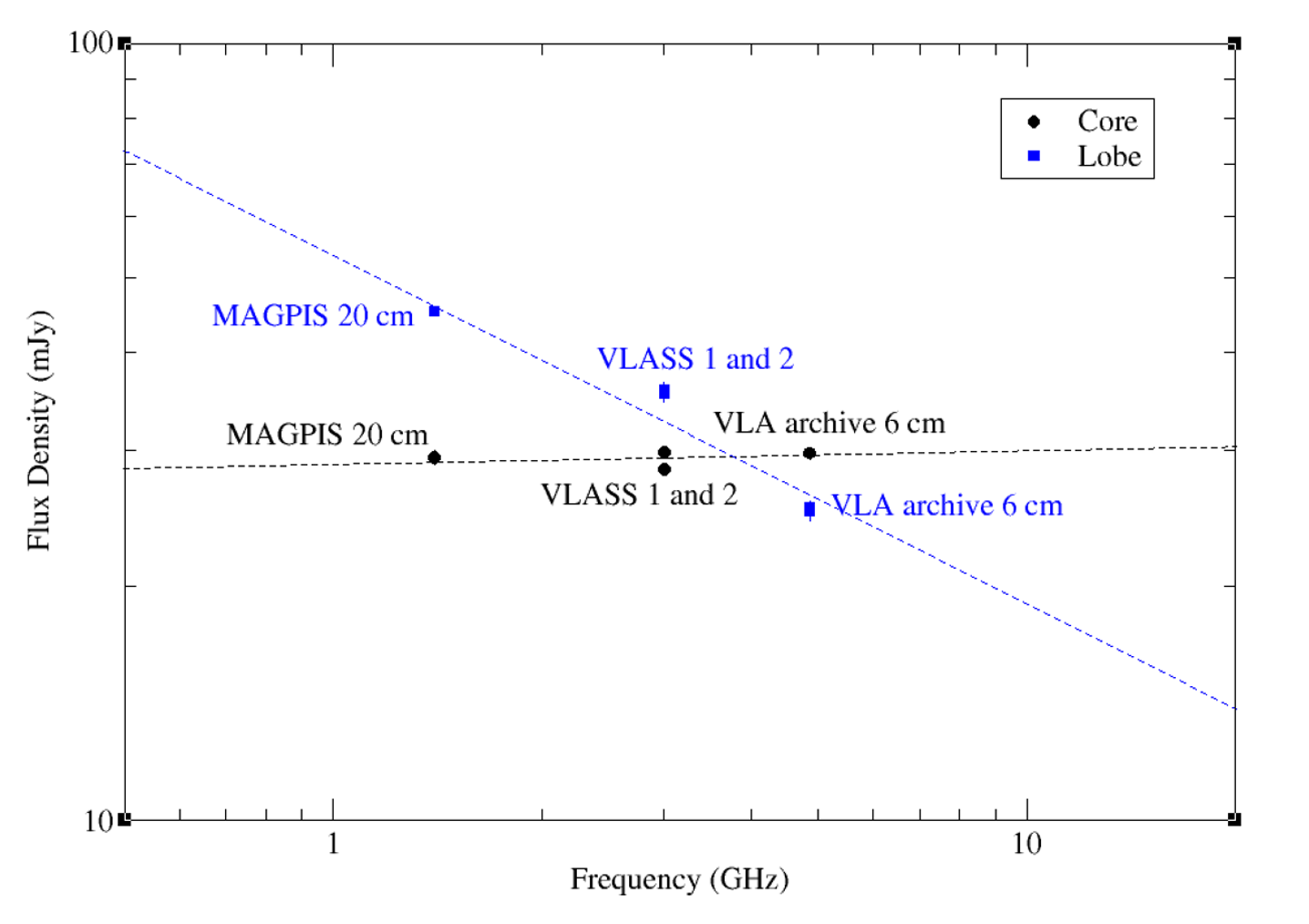}
      \caption{Radio spectra of \iras\ components. The dashed lines represent the least-squares fits to the VLA flux densities  
      in Table \ref{TS2}. The black and blue colours correspond to the central core and to the jet-powered extended lobe, respectively.
      }
         \label{S2}
   \end{figure}

To further investigate the nature of the radio emission, we compiled flux density measurements from archival interferometric observations and public radio surveys with comparable angular resolution, spanning frequencies from $\simeq 1$ to 10 GHz (Table \ref{TS2}). 
All measurements were performed consistently within AIPS using Gaussian fits with its JMFIT task. Primary beam corrections were applied when required (project AB573). Assuming moderate variability in the absence of strictly simultaneous observations, we fitted the radio spectra of the compact core and the extended jet/lobe component independently. 
The core spectrum is well described by a flat power law,   
$S_{\nu} = (28.7 \pm 0.7 {\rm ~mJy}) [\nu/{\rm GHz}]^{+0.02 \pm 0.02}$, while the extended component follows a steeper spectrum, 
$S_{\nu} = (53    \pm 2    {\rm mJy}) [\nu/{\rm GHz}]^{-0.45 \pm 0.03}$ (Fig. \ref{S2}). These spectral indices are consistent with partially self-absorbed synchrotron emission from a compact jet base and optically thin synchrotron radiation upstream from a jet-powered radio lobe.

\begin{table}
\caption{Measurements with best matching angular resolution}
\label{TS2}
\centering
\begin{tabular}{c c c c c}
\hline\hline
Origin 	& Epoch & Freq.  & Core  &  Lobe \\
 of            &           &  (GHz)       & Flux  &  Flux \\
 Data       &          &                  & Density & Density \\
                 &           &                 &(mJy)                      & (mJy)                  \\
\hline 
AB573	        & 1990/12/07 &	4.86	& $29.7 \pm 0.5$  & 	$25.0 \pm 0.7$ \\
	                 &1990/12/09 &	4.86	& $29.7 \pm 0.4$ &	$25.3 \pm 0.5$ \\
MAGPIS           & 2000-2001(*)&1.4 & $29.3 \pm 0.6$ &	$45.3 \pm 0.5$ \\
VLASS1           & 2019/03/10  & 3.0  & $28.3 \pm 0.4$ & 	$35.8 \pm 0.8$ \\
VLASS2           & 2022/01/15  & 3.0   &$29.8 \pm 0.4$ &  $35.5	\pm 0.9$\\
\hline
\end{tabular}
\tablefoot{(*) Multi-epoch combined survey \citep{ref13}.}
\end{table}

\section{Constraints on jet speed and inclination from Doppler asymmetry} 
\label{ratio}

The strong one-sidedness observed in the 6 cm VLA map in Fig. \ref{F1} can be interpreted as relativistic Doppler boosting. 
The peak flux density of the central component reaches up to 8.5 mJy beam$^{-1}$ (uncorrected for primary beam response) 
and exhibits a flat spectral index  $\alpha \simeq 0$. Following an approach similar to \citet{2012A&A...538L...1K} in blazar-like source IC~310, 
we assume it to be representative of the youngest approaching jet, while no trace of the receding jet is detected. Adopting as upper limit three times the map rms noise of 110 $\mu$Jy beam$^{-1}$, the jet-to-counterjet brightness ratio fulfills $R > 26$. 
 Primary beam response plays no role here as it cancels out when taking the ratio. 
 
 For a continuous jet, the velocity $\beta=v/c$  in speed of light units and the line-of-sight inclination angle 
 $\theta$  can be constrained according to  
 $\beta \cos{\theta} > (R^{1/(2-\alpha)} - 1 )/(R^{1/(2-\alpha)} + 1)$.
 This implies $\beta \cos{\theta} >  0.672$, so $\beta > 0.672$  and $\theta \leq 47.8^{\circ}$.
Just as an illustrative example, assuming for instance an intermediate value $\beta= 0.75$ the inferred inclination  would be  $\theta \simeq 26^{\circ}$. 

\section{Milli-arcsecond-scale structure} \label{VLBI} 

We observed \iras\ with the European VLBI Network (EVN) on 2024 May 26 for a total duration of 8 h (project EM178). 
The wavelength of 18 cm (equivalent to a frequency of 1.67 GHz) was selected as a compromise between sensitivity and sharpest point spread function. 
The observations used standard phase-referencing technique (with a cycle of 3 minutes on target and 2 minutes) on the nearby phase calibrator (J1832$-$1035) at a maximum recording bitrate of 1 Gbps divided into four 32 MHz subbands. We also observed a check source (J1828$-$0912) to verify the absolute astrometry and calibration. Data reduction followed standard EVN pipeline processing within AIPS, supplemented by manual flagging of corrupted visibilities. 

Imaging was performed with IMAGR using a zero ROBUST parameter, achieving a rms noise level of $\simeq 130$ $\mu$Jy beam$^{-1}$.
 Clean components were restricted to a central region to minimise spurious extended artifacts. The resulting EVN image (Fig. \ref{F1}, bottom right panel) 
 resolves the compact radio core into an elongated, one-sided structure aligned with the arcsecond-scale collimated emission seen in the VLA data. 
 The peak flux density is at the level of 1.4 mJy beam$^{-1}$. Its radio coordinates, 
 RA = $18^h 32^m 08.9340^s \pm 0.0005^s$, Dec = $-09^{\circ} 39^{\prime} 06.044^{\prime\prime} \pm  0.007^{\prime\prime}$ (ICRS), 
 are consistent within 95\% uncertainty with the {\it Gaia} DR3 optical position16 after accounting for proper motion and phase-reference errors. 
 The absence of detectable counterjet emission, above three times the rms noise, yields a jet-to-counterjet ratio $R > 3.6$ at milli-arcsecond scales. 
 This provides an independent, albeit less restrictive ($\beta > 0.310$ and $\theta \leq 71.9^{\circ}$), confirmation of relativistic beaming. Together with the VLA results, these observations firmly establish the presence of a compact, relativistic jet flow aligned from milli-arcsecond to arcsecond scales.

\section{Optical spectroscopy} \label{caha} 

\begin{figure}
   \centering
   \includegraphics[angle=-0,width=\columnwidth]{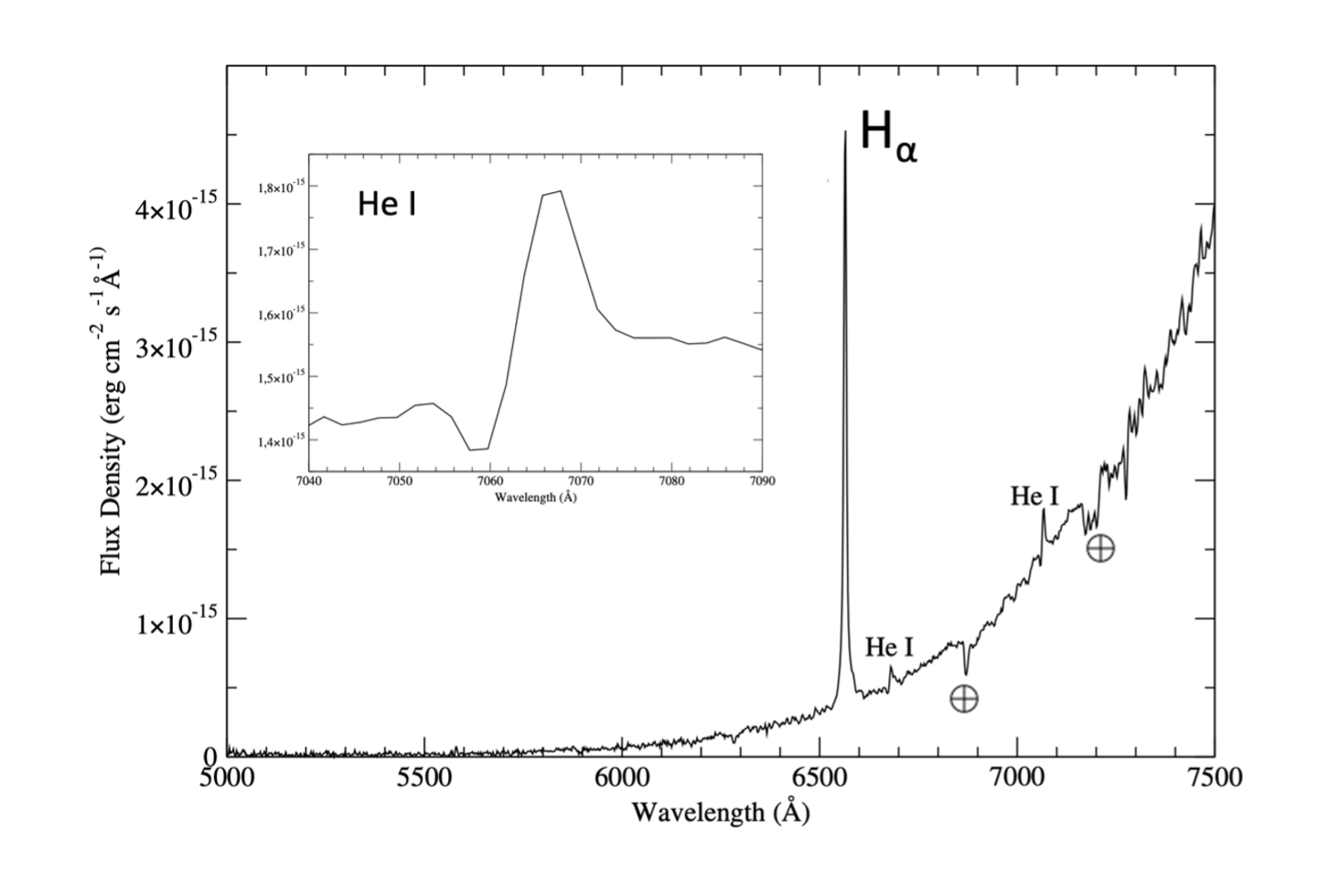}
      \caption{
      CAHA optical spectrum of \iras. H$\alpha$v and He I emission lines are the most prominent features. 
      The inset is a zoom around the He I 7065 \AA\ line illustrating the presence of a P-Cygni profile. Telluric absorption features are marked with an Earth symbol.  
      }
         \label{S3}
   \end{figure}

We obtained optical spectra of \iras\ with the 2.2 m telescope at the Centro Astronómico Hispano Andaluz (CAHA, Calar Alto, Spain) using the CAFOS instrument for classification purposes. Eight nearly daily spectra were acquired between 2025 May 24 and June 2, for a total integration time of $\simeq 9$ h. The median-combined spectrum is shown in Fig. \ref{S3}. Standard long-slit reduction procedures (bias subtraction, flat-fielding, extraction, and wavelength and flux calibration) were applied using IRAF. 
The overall spectral appearance is consistent with historic reports \citep{ref11}. 
In particular, the prominent single-peaked H$\alpha$ emission line, 
emerging above a heavily absorbed continuum, shows a comparable equivalent width of $-120 \pm 10$ \AA. 
 The improved signal-to-noise ratio reveals clear P Cygni profiles in several He I lines, indicating a stellar mass-loss outflow with velocity of
 $\simeq 300$-$400$ km s$^{-1}$. 
  These features confirm previous suspicions about the presence of a dense, dusty circumstellar envelope \citep{ref11}, 
  which is in accordance to the observed warm dust emission (see infrared subsection below).
  
  Owing to the extreme reddening, already pointed out by \citet{ref11} ($E(B-V) \simeq 6$-$7$ mag), the continuum remains compatible with a broad range of spectral types at several kiloparsecs. Nevertheless, the hydrogen and helium emission-line spectrum favours a giant or supergiant early-type (O/B) companion in a high-mass X-ray binary configuration, although a luminous blue variable nature cannot be fully excluded.

\section{Multi-epoch optical photometry of \iras\ and variability}  \label{photo}

   \begin{figure}
   \centering
   \includegraphics[angle=-0,width=\columnwidth]{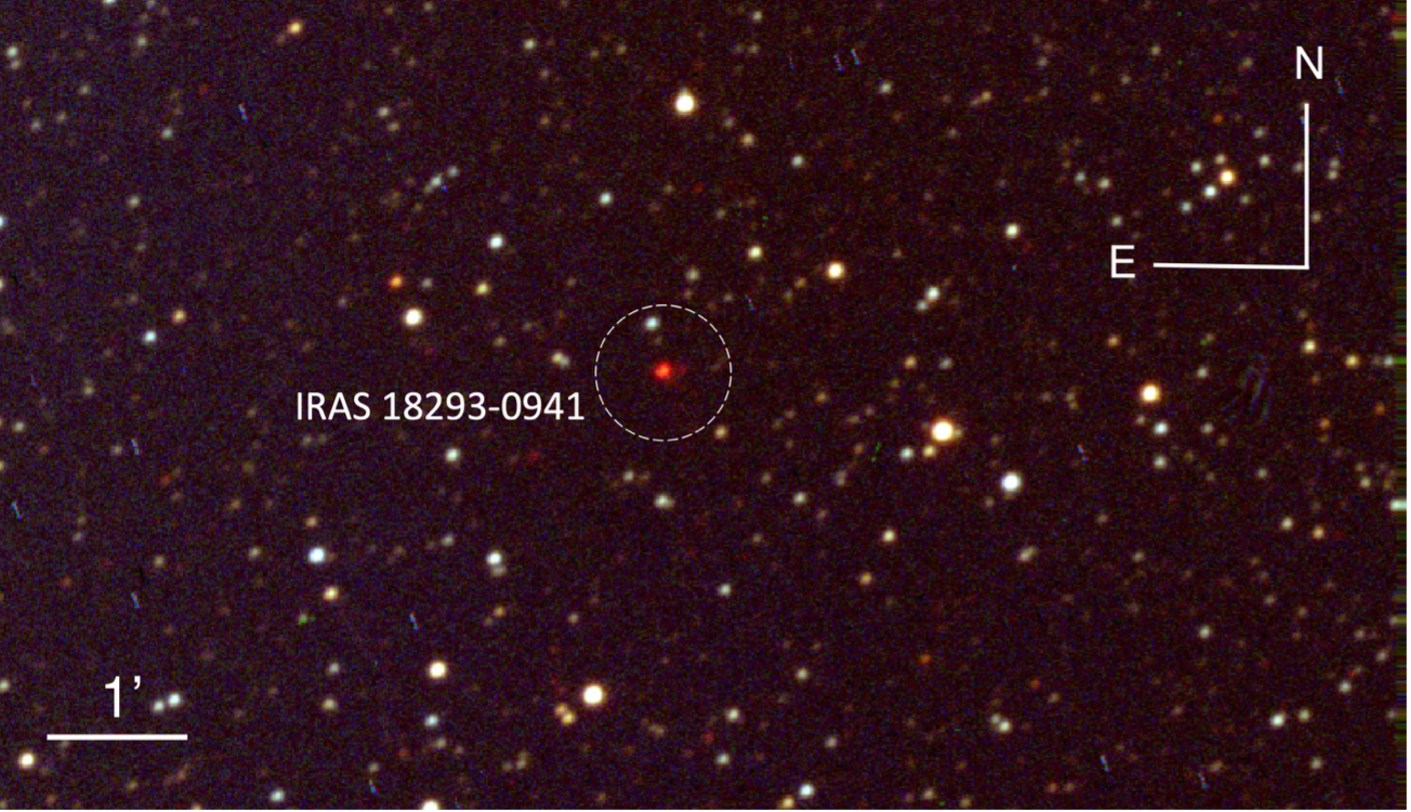}
      \caption{
      Deep trichromatic image of \iras\ obtained with the UJT at the University of Jaén Astronomical Observatory (MPC L83), in 2023 September 30. The red, green and blue layers correspond to the $V$, $R_c$ and $I_c$ filters, respectively. 
            }
         \label{S4}
   \end{figure}

   \begin{figure}
   \centering
   \includegraphics[angle=-0,width=\columnwidth]{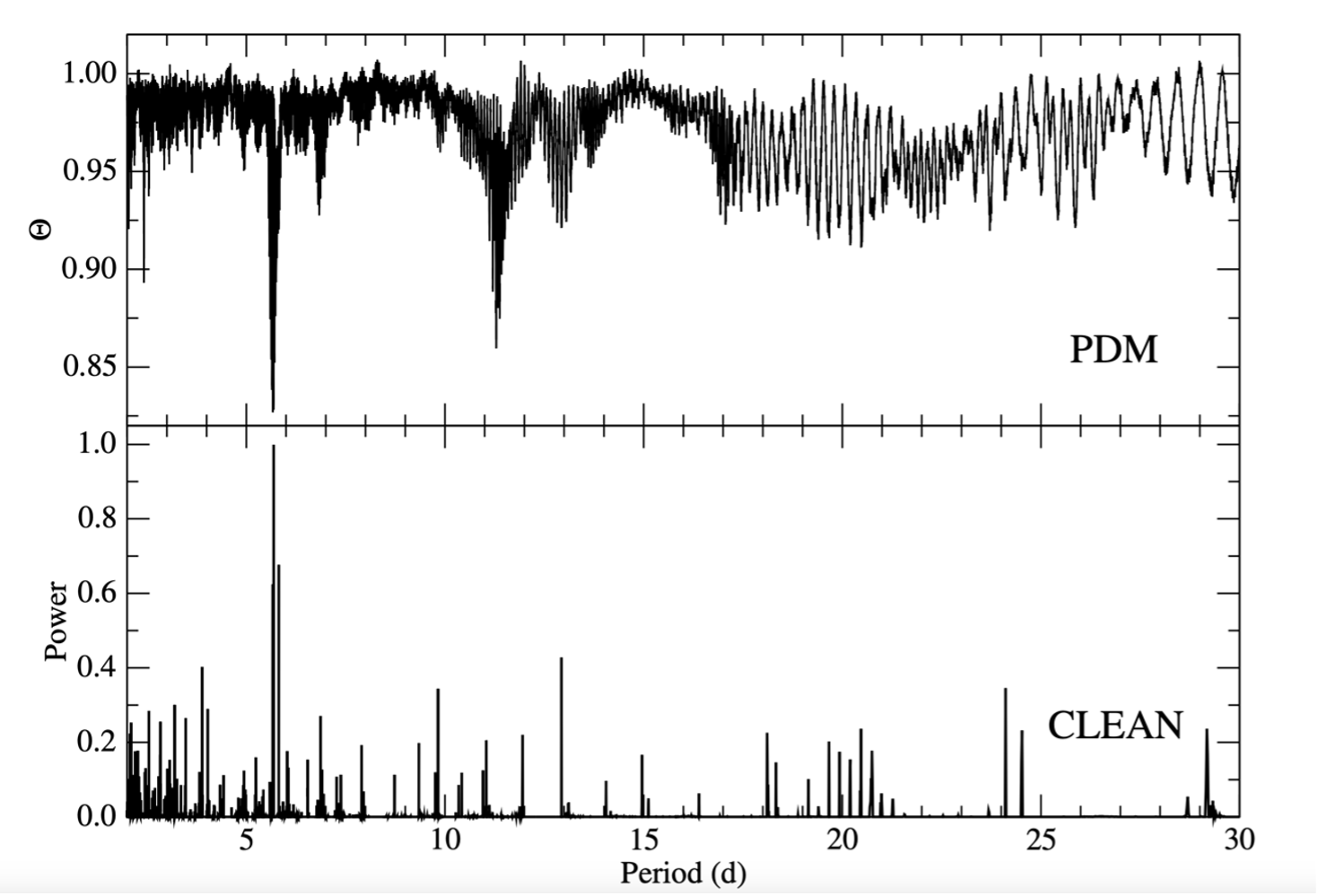}
      \caption{
      Periodograms of $I$-band optical photometry of \iras\ obtained from the combined ZTF and TJO data sets using the PDM and CLEAN algorithms. A previous detrending of the data was applied by subtracting a least-squares parabolic fit.  
      Both methods reveal a prominent peak at $5.688 \pm 0.005$ d that we interpret as half the orbital period.
            }
         \label{S5}
   \end{figure}

   \begin{figure}
   \centering
   \includegraphics[angle=-0,width=\columnwidth]{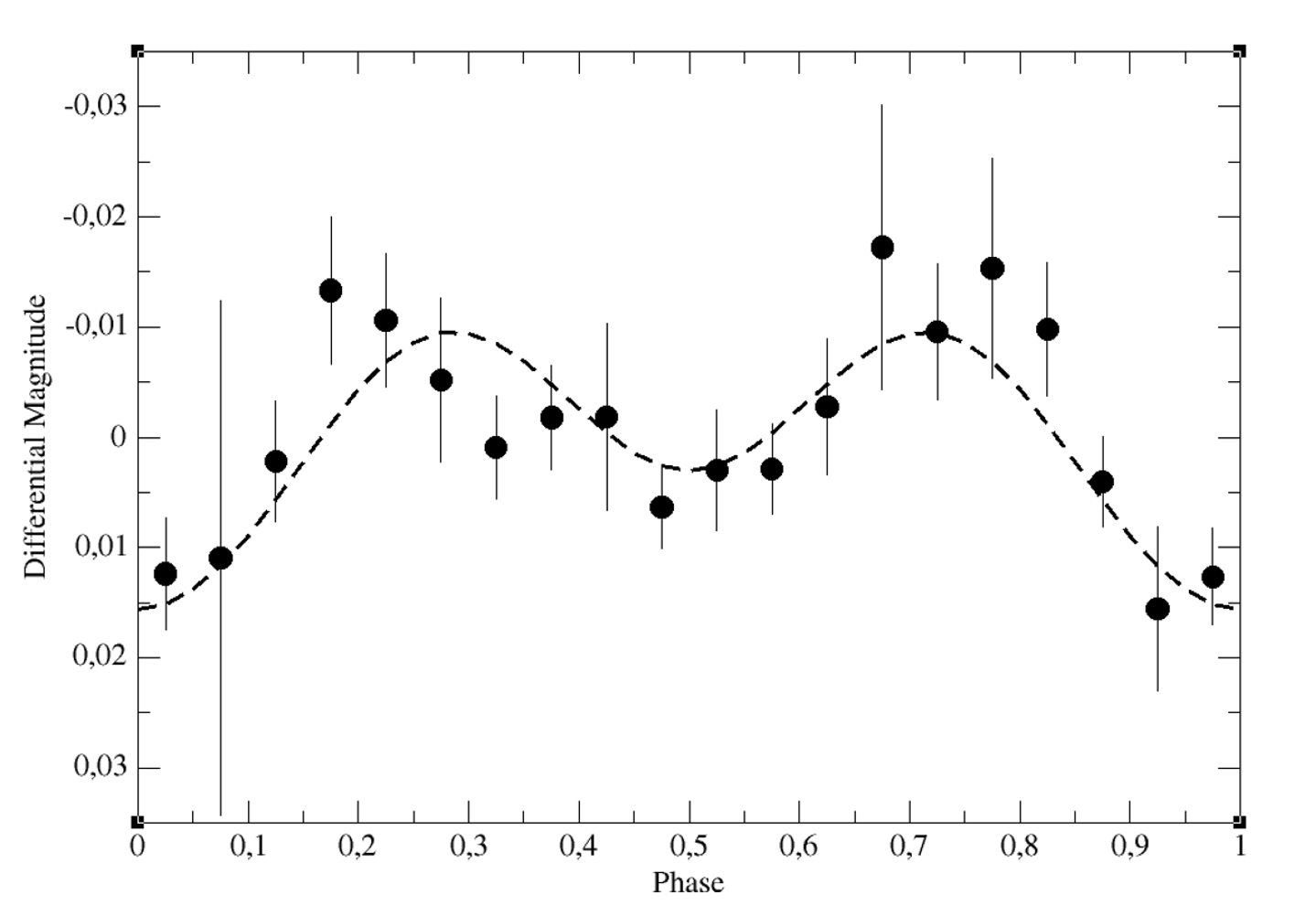}
      \caption{
      Folded and averaged $I$-band light curve of \iras. The orbital period used is $11.38 \pm 0.01$ d 
      and orbital phase zero has been set at MJD 58958.708. Notice the presence of two nearly symmetric maxima, at phases 0.25 and 0.75, as expected from an ellipsoidal variability due to a distorted luminous star. The dashed line is a crude attempt to fit the data based on a simple Wilson and Devinney approach. 
            }
         \label{S6}
   \end{figure}

Time-domain optical monitoring was initiated with the 0.4 m University of Jaén Telescope  \citep{ref36}, 
 which revealed day-scale variability at the $\simeq 0.01$ mag level. This prompted an extended campaign with the 0.8 m Telescope Joan Or\'o (TJO, Montsec Observatory, Spain) between 2024 March and October (90 nights), 
 complemented by 146 nights of Zwicky Transient Facility \citep{ref37}  (ZTF) observations obtained between 2020 April and October. 
 All images were acquired in the $I$-band and analysed using standard differential aperture photometry procedures within IRAF. Because of the extreme reddening, see the field of view in Fig. \ref{S4}, reliable measurements were only possible in this band where the source remains optically bright around 12.3 magnitude. A combined (TJO+ZTF) periodogram analysis using the Phase Dispersion Minimization \citep{ref38}  (PDM) 
 and CLEAN \citep{ref39} algorithms (Fig. \ref{S5}) reveals a dominant period at $5.688 \pm 0.005$ d, independently recovered in  individual datasets. 
 A classical Lomb–Scargle analysis yields a consistent result. Given the typical double-peaked morphology of ellipsoidal variability in binaries, the inferred orbital period doubles to $11.38 \pm 0.01$ d. 
 
 The folded and averaged light curve (Fig. \ref{S6}) shows a very small-amplitude ($\simeq 0.02$ mag) modulation, that would imply a low orbital inclination in an ellipsoidal variability context. Our ignorance about precise stellar masses and effective temperatures renders difficult an accurate modelling at this stage. However, simple Wilson–Devinney–type modelling \citep{ref40} suggests inclination values around $\simeq 20^{\circ}$ for representative OB-star parameters and a few solar masses compact companion (dashed line in Fig. \ref{S6}). Consequently, a similarly reduced jet angle with the line of sight is inferred provided that it emerges perpendicularly to the orbital plane consistent with Appendix \ref{ratio} results. 

Concerning rapid optical variability, if classical optical flickering is present, its amplitude does not exceed $\sim 0.01$ mag. However, this does not contradict the microblazar interpretation.  In \iras\, the optical emission is likely dominated by radiation reprocessed in a dense circumstellar envelope. Outflowing material, will act as an extended re-emitting region and will efficiently damp short-timescale fluctuations originated in the inner accretion flow or jet. This suppression is expected to be particularly strong in the reddish I-band filter that extinction forced us to use.

\section{Large-scale radio structure from MeerKAT imaging}  \label{hospot}

To probe the environment on larger spatial scales, we analysed deep L-band images of the \iras\ 
 field from the SARAO MeerKAT 1.3 GHz Galactic Plane Survey \citep{ref20}. 
 With an angular resolution of 8 arc-second and rms noise level approaching 26 $\mu$Jy beam$^{-1}$ in this field, the deep MeerKAT image reveals extended radio emission aligned with the VLA and EVN jet axes, forming an edge-brightened, 
 elongated structure that traces the boundaries of a jet-inflated cocoon (top left panel in Fig. \ref{F1}). The radio emission is predominantly limb-brightened, with little or no diffuse emission detected in the interior. 
 
 On the Western receding-jet side, the outer boundary shows a localized brightness enhancement consistent with a terminal interaction region (hotspot). In contrast, the Eastern approaching-jet side displays a more diffuse outer rim without a so compact terminal feature. The marked contrast between the well-defined receding hotspot and the more diffuse approaching-side rim reveals a pronounced density gradient in the surrounding medium. This asymmetry is naturally explained if the receding jet is impacting a denser region associated with the nearby molecular cloud, while the approaching jet propagates into a more rarefied environment. This large-scale environmental stratification provides a natural framework for the asymmetric jet morphology and for the efficient particle acceleration inferred at the receding termination region.

\section{Radio spectral-index mapping of the large-scale structure} \label{spix}

We also inspected the SARAO spatially resolved spectral-index maps \citep{ref20}
 derived taking advantage from its wide bandwidth (672 MHz) in order to distinguish between thermal and non-thermal emission components (Fig. \ref{F1}, bottom left panel). The compact core and inner jet exhibit flat-to-moderately steep spectra consistent with synchrotron emission. At larger scales, the prominent Western hotspot in the inner rim of the receding cocoon shows a clearly non-thermal spectral index ($\alpha=-0.7\pm 0.1$), coincident with the MeerKAT brightness enhancement and consistent with synchrotron emission from shock-accelerated particles. A localized region of non-thermal emission is also detected along the Eastern extrapolated direction of the approaching jet. Although no compact hotspot is present, such feature points to the existence of a less developed bow-shock also there. This suggests also ongoing particle acceleration despite the lower confinement efficiency on that side.

\section{Optical shock tracers along the large-scale structure}  \label{shocks}

We inspected narrow-band optical data from the VST Photometric H$\alpha$ Survey of the Southern Galactic Plane (VPHAS+) \citep{ref23}
 to search for ionized gas associated with the radio structures. The VPHAS+ images reveal faint extended H$\alpha$ emission outlining portions of the radio cocoon boundary. In addition, wide-field narrow-band imaging from the Northern Sky Narrowband Survey (DR0.2) \citep{ref24}
  also shows H$\alpha$ filamentary emission spatially coincident with the outer rim of the receding-side cocoon and with the localized radio hotspot (Fig. \ref{F2}). The combined morphology is suggestive of shock-excited gas at the interface between the relativistic outflow and the ambient medium. Interestingly and more than a half century ago \citep{ref41}, the compact emission nebula [GS55]150 was catalogued just at the position of the receding hotspot as an uncategorized ‘interstellar medium object’.  No similar 
  nebulous counterpart appears 
  at the approaching side.

\section{Far-infrared signature of jet-dust interaction} \label{Herchel}

We retrieved archival far-infrared observations of the \iras\ field obtained with the Herschel Space Observatory \citep{ref19}. 
PACS maps at 70 $\mu$m were analysed as they provide optimal sensitivity to warm dust. As shown in Fig. 2, the 70 $\mu$m 
emission reveals a shell-like structure that is remarkably matching the 6 cm radio contours from archive VLA observations. 
The spatial correlation between the radio jet morphology and the dust emission peak suggests that the relativistic outflow is actively interacting with and sweeping up the surrounding interstellar medium at relatively small distances from the binary. This observational evidence supports the presence of the dusty shell modeled in our theoretical scenario below. In addition, the far-infrared emission traces structures spatially coincident with the radio morphology at larger scales. Enhanced 70 $\mu$m emission is detected at the position of the receding hotspot and along portions of the cocoon boundary, indicating the presence of warm dust mixed with the shocked gas. Fainter far-infrared emission is also present toward the approaching side, although without a compact radio counterpart. The coexistence of non-thermal radio emission and warm dust at the cocoon boundary suggests that a fraction of the dust survives the passage of the jet-driven shock and becomes compressed and heated in the post-shock region.

\section{Distance estimate}   \label{dist}

The distance to \iras\ is not directly constrained by {\it Gaia} astrometry, which reports a formally negative parallax \citep{ref16}. 
Early work \citep{ref11}  placed the system at distances of order $\simeq 8$-$9$ kpc based primarily on its extreme optical reddening. 
However, our multi-wavelength analysis indicates that a significant fraction of the extinction is intrinsic to the system, likely arising in the dense circumstellar environment described in this work. Based on CO emission line surveys \citep{ref42}, a molecular cloud with Galactic coordinates (21.97, $-$0.29) has been identified at a kinematic distance of $\simeq 3.6$ kpc \citep{ref25}. As visible in Fig. \ref{F3}, its brightest and densest peak is spatially coincident with the receding-jet termination region. Given the clear morphological evidence for jet–cloud interaction presented in the main text, we adopt this value as the fiducial distance to \iras\ throughout this work. We note that moderate deviations from this distance would not qualitatively affect our main conclusions regarding jet orientation, environmental interaction, or particle acceleration efficiency.

\section{X-ray observations with XMM-Newton and Chandra}  \label{xray}

\begin{table}
\caption{Best fit parameters and $1 \sigma$ level uncertainties derived from fitting both XMM-Newton observations with and absorbed pshock model}
\label{TS3}
\centering
\begin{tabular}{l c c}
\hline\hline
Parameter	 & Obs. 0673720201	& Obs. 0673720401\\
\hline
Constant   &	$1.08 \pm 0.03$ & $1.09 \pm 0.03$ \\
(MOS1)    &                               &                           \\
\hline
Constant    &   $1.08 \pm 0.03$ &  $1.01 \pm 0.03$ \\
(MOS2)	&                                &                            \\
\hline
$N_H$      &  $0.9 \pm 0.1$       & $1.1 \pm 0.1$ \\ 
($10^{22}$  cm$^{-2}$)	&   & \\
\hline
$kT$       & $4.0 \pm 0.2$  & $3.95 \pm 0.15$\\ 
(keV) & & \\
\hline
Abundance & $0.31 \pm 0.03$ & $0.27 \pm 0.03$\\
(Solar)  &  &  \\
\hline
Ionisation  &  &  \\
 timescale & $2.3^{+0.7}_{-0.5}$ & $1.6^{+0.4}_{-0.3}$ \\
 ($10^{11}$ s cm$^{-3}$)	&  & \\
 \hline
Unabsorbed flux & $0.85 \pm 0.07$ & $1.05 \pm 0.1$\\ 
($10^{-11}$ erg s$^{-1}$ cm$^{-2}$)   &  & \\
0.5-10 keV  &  & \\
\hline
$\xi^2$ /  d.o.f. &	244.6 / 222  &	247.8 / 222\\
\hline
\end{tabular}
\end{table}

\begin{figure}
   \centering
   \includegraphics[angle=-0,width=\columnwidth]{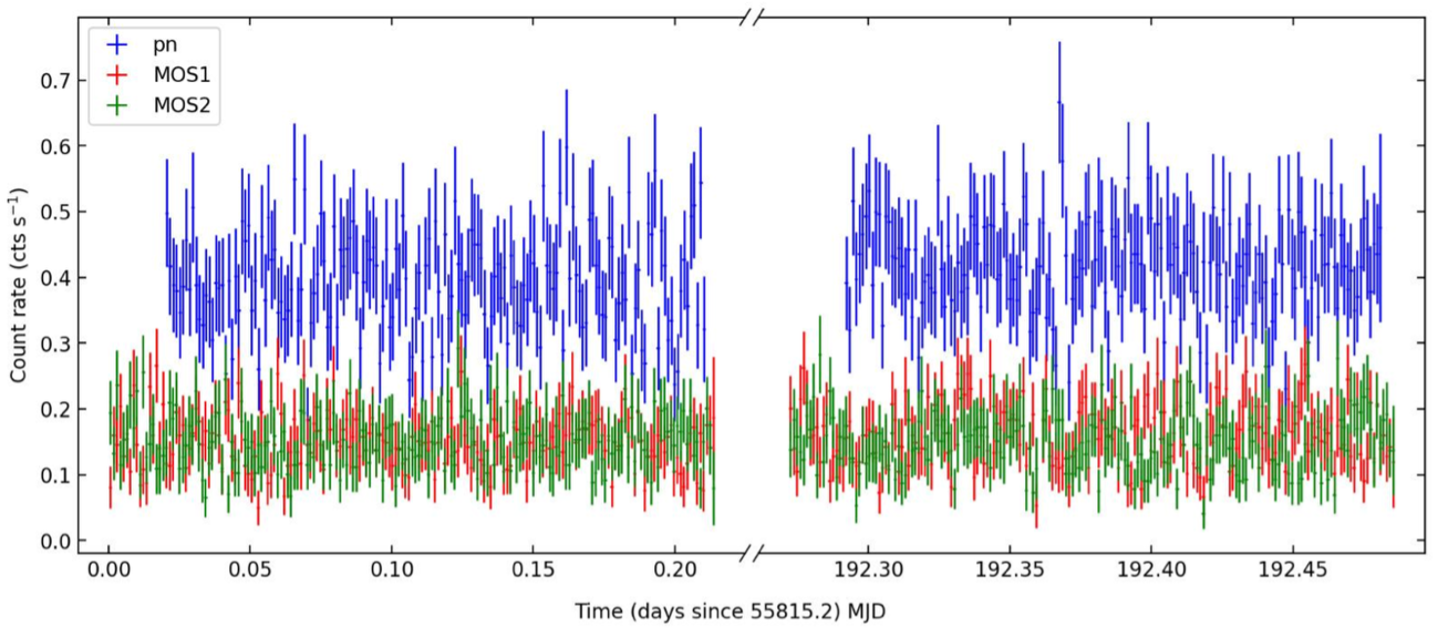}
      \caption{
      EPIC pn, MOS1 and MOS2 background subtracted light curves of \iras. Extraction was in the 0.5 keV to 10 keV energy range, using a bin size of 100 seconds. Both XMM-Newton observations are shown on the same plot, separated by the split time axis.
                  }
         \label{S7}
   \end{figure}

Two 18 ks pointed observations of \iras\ are available in the XMM-Newton archives (ObsIDs 0673720201 and 0673720401) carried out on 2011 September 11 and 2012 March 21, respectively. The MOS and pn cameras were in Large Window Mode in both of them. The data were reduced using the v21.0.0 version of the XMM-Newton Science Analysis System (SAS) software and up-to-date CCF calibration files as of 2024 October.  Spectra were modelled using XSPEC \citep{ref43}.
Fluxes and uncertainties ($1 \sigma$ ) were extracted from MCMC realisations, using the chain and cflux commands from XSPEC.  
The background subtracted light curves were extracted with a bin size of 100 s in the 0.5-10 keV energy range (see Fig. \ref{S7}). 

\begin{figure}
   \centering
   \includegraphics[angle=-0,width=\columnwidth]{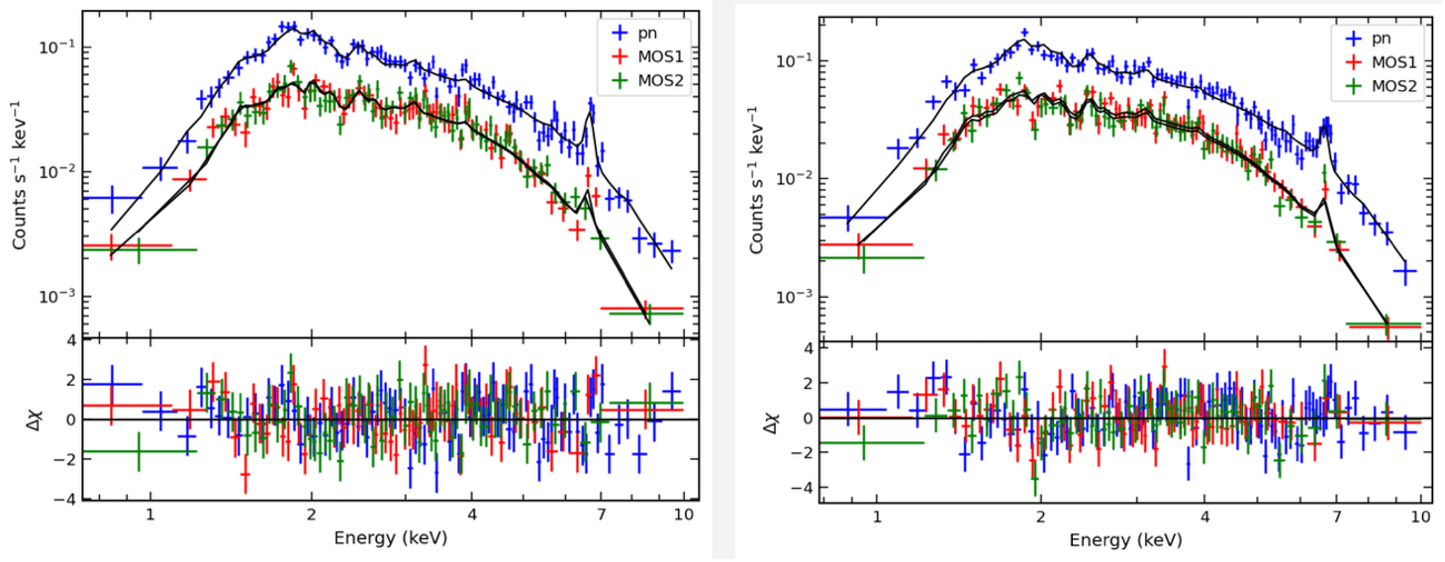}
      \caption{
      EPIC pn, MOS1 and MOS2 time-averaged, background subtracted spectra of \iras. On the left and right panels we show observations 0673720201
      and 0673720401, respectively.  Spectra are fitted with an absorbed pshock model. Residuals are shown on the bottom panel of each plot. 
                  }
         \label{S8}
   \end{figure}

\iras\  shows no apparent variability in its count rate in the time scale of hours between both observations. We extracted power spectra for both observations using the powspec task from HEASoft, using 10s bin light curves in the 0.5 keV to 10 keV energy band. No pulsations are significantly detected above Poisson noise in the 0.0005 to 0.05 Hz frequency range. We also fitted the time average spectra with an absorbed pshock model, which describes the emission from a shocked plasma in thermal equilibrium. The time averaged spectra, best-fitting models and residuals of each observation are shown in Fig. \ref{S8}. A summary of the best fits to both observations is presented in Table \ref{TS3}. It is clear from this table that all model parameters are consistent within uncertainties in between both observations. We obtain very good fits in both cases, with reduced $\chi^2$ of approximately 1.1 for 222 degrees of freedom. The spectra of both observations can be described by a heavily absorbed ($N_H \sim 1 \times 10^{22}$ cm$^{-2}$) plasma of temperature close to 4 keV and subsolar abundances ($A \sim 0.3$). We obtained absorption corrected fluxes of approximately $9 \times 10^{-12}$ erg s$^{-1}$ cm$^{-2}$ in both observing epochs in the 0.5 to 10 keV energy range, which translates into an unabsorbed X-ray luminosity at the assumed distance of 3.6 kpc of $L_X$ (0.5$-$10 keV) $ \approx 1.4 \times 10^{34}$ erg s$^{-1}$. 
The lack of strong short-term variability and the moderate luminosity are consistent with thermal emission produced in shocks formed by colliding winds originated in the accretion disk and the companion star.

The derived X-ray power renders \iras\ apparently over-luminous in radio by nearly two orders of magnitude with respect to sub-Eddington accreting systems where
a strong radio/X-ray correlation holds \citep{2003MNRAS.345.1057M}, thus suggesting a significantly different physical scenario (Appendix \ref{model}). 
We end noting that a pointing on our target also exists in the Chandra archive obtained on 2008 May 31 (ObsID 9611). However, the exposure time was so short (1 ks), that we only used it to verify the X-ray position agreement with the optical and radio coordinates within less than one arcsecond. 

\section{Gamma-ray sources in the \iras\ field}  \label{arnau}

Figure \ref{F3}  summarizes the gamma-ray population in the region against a background of CO molecular line emission \citep{ref42}. 
We first analysed Fermi Large Area Telescope (LAT) archival data for the crowded Galactic-plane field surrounding \iras.
 In the last Fermi-LAT catalog \citep{ref34}, there is only a single GeV source (4FGL J1830.8$-$0947) close to our target, although several pulsars are present that could potentially contaminate the background. In particular two of them, PSR J1831$-$0952 and PSR J1833$-$1034, are relatively bright and located within one degree of the IRAS source. To remove their effect, we applied pulsar gating using the available timing ephemerides. The total LAT exposure with valid ephemerides for both pulsars amounts to 9.33 years, yielding an effective exposure of 5.67 years after simultaneous gating. As shown in Fig. \ref{F3}, the emission of 4FGL J1830.8$-$0947 survives the pulsar gating and remains significant. The revised confidence ellipse is displayed relative to the 4FGL-DR4 position, showing an offset of $0.25^{\circ}$. 
 For reference, Fig. \ref{F3} also marks the location of relevant UHE and VHE sources in the field: 1LHAASO J1831$-$1007u$^*$ \citep{ref10}, 
 HESS J1831$-$098, and 3HWC J1831$-$095 \citep{ref44}. The asterisk in the LHAASO designation indicates potential source confusion in the region, consistent with the presence of multiple nearby pulsars. The molecular cloud (21.97, $-0.29$) \citep{ref25}, proposed as the target of the \iras\ jets, accounts for most of the CO background. We notice that our revised location of 4FGL J1830.8$-$0947 excludes the IRAS source and is offset with respect to the molecular cloud. Therefore, we could be dealing with an unrelated Fermi-LAT emitter. Finally, spectral upper limits at the IRAS position were derived to better constrain our physical understanding.

\section{Theoretical modelling} \label{model}

We model the system assuming a binary embedded within a dusty circumstellar shell. Winds from the supercritical accretion disk and the massive companion star collide, forming a shocked interaction region as inferred from X-ray spectra. Two bipolar jets emerge from the binary and propagate in opposite directions into the surrounding interstellar medium, breaking through the shell. A nearby molecular cloud is also present along the jet axis. The adopted distance to the system is that of the molecular cloud kinematically derived through its CO spectral line \citep{ref25}, $d = 3.6$ kpc. 
The broadband emission observed toward the source can be reproduced by the combined contribution of the colliding-wind region, the dusty shell, the relativistic jets, and a nearby cloud. Below we describe these components, which together define the schematic scenario shown in the top panel of Fig. \ref{F4}. The bottom panel in the same figure presents their respective contribution to the SED of the microblazar-molecular cloud system. Model parameters are listed in Table \ref{TS4}.
Their adoption is justified to provide a consistent fit from radio to the PeV domain based on previous theoretical works
\citep{ref45,ref29, ref35}.

\subsection{Binary system} 

We assume a black hole of mass 10 $M_{\odot}$ accreting from a massive companion star at super-Eddington rates, 
$\dot{M} = 2.2 \times 10^{-6}$ $M_{\odot}$ yr$^{-1}$, The stellar radius and temperature are taken as 
$R_{\rm *} = 2 \times 10^{12}$ cm and $T_{\rm *} = 4.5 \times 10^4$ K. In the supercritical regime, the accretion flow self-regulates near the Eddington rate while expelling excess material through a powerful optically thick wind. This wind absorbs the X-ray emission produced in the inner disk and reprocesses it to lower frequencies from its photosphere. Therefore, the wind blocks the X-ray emission from the disk from the perspective of an external observer. For the adopted accretion rate, the photospheric height is $\sim 10^9$ cm \citep{ref29}. 

\subsection{Colliding winds}

 The disk-driven wind interacts with the stellar wind, producing a double-shock structure \citep{ref45}.  
 One of the shocks is expected to be radiative and to generate thermal emission.  We model this component as a modified blackbody out of strict equilibrium, representing a hot plasma region with characteristic temperature $T \sim 10^7$ K that cools as it propagates away from the interaction site. The integrated emission from this region accounts for the observed X-ray flux. 

\subsection{Shell}

An opaque dusty shell surrounds the binary system. We describe it using two layers: an inner thin region and an outer thick envelope. The inner layer absorbs ultraviolet radiation from the star, which is subsequently reprocessed and thermally re-emitted by the outer shell. We model both components as modified blackbodies corresponding to opaque dust in the inner layer ($T_{\rm in} = 2400$ K) and more transparent dust in the outer envelope ($T_{\rm out} = 45$ K).

\subsection{Jet} 

Guided by the observed radio morphology and Doppler-boosting constraints previously derived, 
we adopt a jet kinetic power corresponding to a fraction $\sim 0.1$ of the accretion power at the magnetization radius \citep{ref29},
 yielding $L_{\rm j} \approx 2 \times 10^{39}$ erg s$^{-1}$. The jet propagates with a velocity of $0.75c$
  and a semi-opening angle of $\approx 6^{\circ}$. Its lateral expansion halts at the height where recollimation and a shock occurs, after which the flow continues with approximately constant radius until reaching a maximum length determined by the source age. 
  Assuming an age $t_{\rm age} = 5  \times 10^4$ yr gives a characteristic jet length $l_{\rm j} \approx 10^{20}$ cm  comparable to the observed binary to hotspot separation.  An adiabatic reverse shock is expected to form at the jet termination region. We describe next the recollimation and reverse shocks. The adopted jet and shock parameters represent plausible time-averaged values over the lifetime of the system and do not necessarily correspond to the present-day lobe dimensions which could be the result of recent episodic activity or a more extended but undetected dark jet. 
  
 \subsection{Recollimation shock} 
  
  Recollimation is assumed to occur when the jet interacts with the dusty shell at a height $3 \times 10^{14}$ cm above the black hole. 
  Particles accelerated to relativistic energies in this region interact with local magnetic and radiation fields. 
  The magnetic field along the jet is parametrized as $B(z)=B_0(z^{\prime}/z)$,
   with $B_0 = 2 \times 10^6$ G at height $z^{\prime}$ near the black hole \citep{ref46}.  
   Stellar photons provide targets for inverse Compton scattering, while the matter density decreases with distance \citep{ref29}. 
   Synchrotron and inverse Compton processes both contribute significantly to the SED. 
   Doppler boosting is included in the observer frame and this component accounts for the radio emission from the compact core.

\subsection{Reverse shock}

To account for the highest-energy emission, 
we assume a strongly amplified magnetic field at the reverse shock in the jet termination \citep{ref47},
 adopting a representative value $B = 100$ $\mu$G. The ambient density is assumed to be $n = 0.1$ cm$^{-3}$, 
  and the shell provides the dominant radiation field. Under these conditions, synchrotron emission from accelerated particles dominates the contribution of this region to the broadband spectrum and reproduces the radio emission observed from the large-scale lobes whose extension in the past lifetime of the system could be wider than today. 

\subsection{Cloud} 

We assume a molecular cloud with characteristic number density $\sim 50$ cm$^{-3}$ and  radius of $\sim 40$ pc roughly matching the angular size of the LHAASO/cloud emission at the adopted distance. The microquasar is offset by $\sim 60$ pc from the cloud centre, 
as suggested by the $\approx$20 arcmin separation from the LHAASO KM2A position assuming a projection angle of  $\approx 20^{\circ}$. 
We calculate the propagation of cosmic rays escaping from the reverse shock of the jet in an irradiation-cloud scenario (48,49) with continuous injection. The diffusion coefficient in the interstellar medium is parameterized as 
$D(E)=D_0 E^{\delta}$,  with $\delta =0.5$ and $D_0 = D(10~{\rm GeV}) = 10^{26}$ cm$^2$ s$^{-1}$. 
Proton–proton interactions between ultra-relativistic particles from the jet and the dense, cold cloud material account for the observed VHE/UHE emission. We also compute synchrotron and Bremsstrahlung radiation from secondary pairs produced in charged-pion decay.

\begin{table} 
\begin{center}
\caption{Main parameters of \iras\ physical model}
\label{TS4}
\begin{adjustbox}{max width=\columnwidth}
\begin{tabular}{l c c c}
\hline
\hline
\rule{0pt}{2.5ex}Parameter & Symbol & Value & Units  \\
\hline
\rule{0pt}{2.5ex}Distance to the source$^{(3)}$ & $d$ & 3.6 & ${\rm kpc}$ \\
Age of the microquasar$^{(1)}$ & $t_{\rm age}$ & $5\times10^{4}$ & ${\rm yr}$ \\
Inclination$^{(3)}$ & $i$  & $20$ & deg \\

\hline
\rule{0pt}{2.5ex}Black hole and accretion disk \\  
\hline 

\rule{0pt}{2.5ex}Black hole mass$^{(1)}$ & $M_{\rm BH}$    & 10 & $M_{\odot}$ \\
Mass accretion rate$^{(1)}$ & $\dot{M}_{\rm input}$ & $2.2\times 10^{-6}$ & $M_{\odot} \ \rm{yr}^{-1}$ \\
Photosphere height of the  & $z_{\rm photo}$ & $2.6\times10^{9}$ & ${\rm cm}$ \\
disk-driven wind$^{(2)}$ & &  &  \\
\hline
\rule{0pt}{2.5ex}Star\\  
\hline
\rule{0pt}{2.5ex}
Effective temperature$^{(1)}$ & $T_{\rm *}$    & $4.5\times 10^4$ & K \\
Radius$^{(1)}$ & $R_{\rm *}$   & $27$  & $R_{\odot}$ \\
\hline
\rule{0pt}{2.5ex}Dusty shell\\  
\hline
\rule{0pt}{2.5ex}Temperature (external layer)$^{(3)}$ & $T_{\rm out}$    & 45 & K \\
Temperature (inner layer)$^{(3)}$ & $T_{\rm in}$    & 2400 & K \\
\hline
\rule{0pt}{2.5ex}Jet \\ 
\hline
\rule{0pt}{2.5ex}Mechanical power$^{(2)}$ & $L_{\rm j}$  & $2\times10^{39}$ & ${\rm erg\,{s}^{-1}}$\\
Semi opening angle$^{(1)}$ & $\theta_{\rm j}$  & $6$  & deg \\
Lorentz factor$^{(3)}$ & $\gamma_{\rm j}$  & $1.51$ & \\
Length$^{(2)}$ & $l_{\rm j}$  & $10^{20}$ & ${\rm cm}$  \\
\hline
\rule{0pt}{2.5ex}Jet recollimation shock \\ 
\hline
\rule{0pt}{2.5ex}Location of acceleration region$^{(2)}$ & $z_{\rm recoll}$  & $3\times10^{14}$ & ${\rm cm}$\\
Magnetic field at  & $B_{\rm recoll}$   & $0.1$ & ${\rm G}$  \\
acceleration point$^{(2)}$ &    &  &  \\
\hline
\rule{0pt}{2.5ex}Jet terminal region\\ 
\hline
\rule{0pt}{2.5ex}Magnetic field at reverse shock$^{(1)}$ & $B_{\rm rs}$   & $100$ & $\mu{\rm G}$  \\
Reverse shock velocity$^{(2)}$ & $v_{\rm rs}$  & $2.25\times10^{10}$ & ${\rm cm\,{s}^{-1}}$\\
\hline
\rule{0pt}{2.5ex}Cloud\\ 
\hline
\rule{0pt}{2.5ex}Radius$^{(3)}$ & $R_{\rm cloud}$  & $40$ & pc\\
Number density$^{(3)}$ & $n_{\rm cloud}$  & $50$  & ${\rm cm^{-3}}$\\
Distance from the microquasar$^{(2)}$ & $D$  & $60$  & ${\rm pc}$\\

\hline
\end{tabular}
\end{adjustbox}
\end{center}
\footnotesize{\textbf{Notes.} We indicate the parameters we have assumed with superscript ${(1)}$, those we have derived  with ${(2)}$, and those taken from observations with $(3)$.}
\end{table}

\subsection{Results}  

The bottom panel in Fig. \ref{F4} shows the spectral energy distributions of the different system components (solid lines) merged together with the observational data (points and dashed lines). The radio emission is reproduced by synchrotron radiation from the jet recollimation and reverse shocks. Infrared emission arises from the two-layer dusty shell, while the $N_H$ surviving X-ray flux is explained by thermal radiation from the colliding-wind region. The VHE component originates primarily from inverse Compton emission at the recollimation shock, and the UHE emission is reproduced through pion decay following proton–proton interactions between jet-accelerated cosmic rays and the nearby molecular cloud.

\clearpage

\end{appendix}


\end{document}